# Evaluating Input Modalities for Pilot-Centered Taxiway Navigation:

## Insights from a Wizard-of-Oz Simulation

1st Chan Chea Mean
*Air Traffic Management Research Institute*
*Nanyang Technological University*
Singapore, Singapore
chan1100@e.ntu.edu.sg

2nd Sameer Alam
*Air Traffic Management Research Institute*
*Nanyang Technological University*
Singapore, Singapore
sameeralam@ntu.edu.sg

3rd Katherine Fennedy
*Air Traffic Management Research Institute*
*Nanyang Technological University*
Singapore, Singapore
katherine.fennedy@ntu.edu.sg

4th Meng-Hsueh Hsieh
*Air Traffic Management Research Institute*
*Nanyang Technological University*
Singapore, Singapore
menghsueh.hsieh@ntu.edu.sg

5th Shiwei Xin
*Air Traffic Management Research Institute*
*Nanyang Technological University*
Singapore, Singapore
xins0002@e.ntu.edu.sg

6th Brian Hilburn
*Center for Human Performance Research*
Philadelphia, United States
brian.hilburn@chpr-usa.com

*Abstract*—Runway and taxiway incursions continue to challenge aviation safety, as pilots often experience disorientation from poor visibility in adverse conditions and cognitive workload in complex airport layouts. Current tools, such as airport moving maps on portable tablets, allow manual route planning but do not dynamically adapt to air traffic controllers' (ATCOs) clearances, limiting their effectiveness in high-stress scenarios. This study investigates the impact of different input modalities—paper-based, keyboard touch, map touch, and speech-to-text—on taxiway navigation performance, using a medium-fidelity flight simulator and a Wizard-of-Oz methodology to simulate ideal automation conditions. Contrary to common assumptions, recent studies indicate that paper-based methods outperform digital counterparts in accuracy and efficiency under certain conditions, highlighting critical limitations in current automation strategies. In response, our study investigates why manual methods may excel and how future automation can be optimized for pilot-centered operations. Employing a Wizard-of-Oz approach, we replicated the full taxiing process—from receiving ATCO clearances to executing maneuvers—and differentiated between readback and execution accuracy. Findings reveal that speech-based systems suffer from low pilot trust, necessitating hybrid solutions that integrate error correction and confidence indicators. These insights contribute to the development of future pilot-centered taxiway assistance that enhance situational awareness, minimize workload, and improve overall operational safety.

## I. INTRODUCTION

Runway and taxiway incursions persist as major safety challenges in aviation, contributing significantly to the overall risk profile of airport ground operations. Despite advancements in navigation tools and procedures, pilots struggle to maintain situational awareness, especially under conditions of poor visibility, unfamiliar airport layouts, and high workload [1].

Existing navigation aids, such as airport moving maps displayed on Electronic Flight Bag (EFB), allow pilots to manually plan their taxi routes. However, these tools often lack the adaptability required to dynamically integrate air traffic controllers' (ATCOs) clearances in real time. This limitation can result in increased workload and potential deviations from intended taxi routes, particularly in high-stress or time-sensitive scenarios [2].

The need for adaptable, pilot-centered automation in aviation is underscored by the foundational work of Endsley in Artman's study [3], [4]. While digital tools are often presumed to outperform traditional methods, recent findings challenge this assumption, suggesting that manual methods, such as paper-based navigation, may offer superior performance in specific contexts [3], [5]. These insights necessitate a critical reassessment of the interplay between human factors and automation to enhance pilot support during ground maneuvers.

Our study builds upon the work of Estes et al. [5], who conducted a systematic evaluation of input modalities for taxiway navigation. Estes et al.'s study, referred to as Evaluation 1, was performed in a desktop setting, potentially limiting its ecological validity. To further advance this research, we conducted our experiment in a medium-fidelity flight simulator, which aims to better replicate real-world pilot interactions and cognitive demands [6]. Additionally, while Estes et al. implemented a functional speech-to-text system as one of their input modalities, their system exhibited a significant error rate of approximately 41%. In contrast, we employed a Wizard-of-

Oz method [1], which allowed us to simulate a perfect recognition system and focus on evaluating pilot performance under ideal speech recognition conditions.

Another key limitation of Estes et al.'s experiment was its use of a single airport environment, which raised concerns about potential learning effects from repeated taxi clearances. To mitigate this issue, our study introduced four custom airport layouts, each featuring modifications designed to alter the visual structure while maintaining equivalent complexity. This ensured that no pilot had prior familiarity with any of the airports. Each input modality was tested on a different airport layout, effectively eliminating any learning effect associated with repeated exposure to the same airport. Furthermore, we employed a counterbalancing approach to systematically vary the sequence of scenarios and input modalities across participants, mitigating potential order effects and ensuring a more reliable assessment of performance differences.

By addressing these limitations and refining the experimental methodology, our study aims to provide a more comprehensive understanding of how different input modalities—paper-based, keyboard touch, map touch, and speech-to-text affect pilots' taxiing navigation performance. The aims of the research work are as follows:

- Evaluation of qualitative pilot responses to different input modalities, informing the future development of Taxiway Navigation Assistance tools
- Identification of operational challenges in current taxi and communication procedures, providing insights into system limitations and areas for improvement

This study addressed the following research questions:

- How do different input modalities affect pilot performance for taxiway clearance and navigation?
- What are pilots' preferences and operational needs for an automated taxiway assistance?

The paper is organized as follows. First, Section II discusses related work and previous research gaps. Section III presents the methodology, including hard and softwares and data collection procedures. Section IV presents both the quantitative and qualitative results, followed by Section V, where the results are discussed and finally the limitation and future work followed by the conclusion are drawn in Section VI and Section VII respectively.

## II. RELATED WORK

### A. Limitations of Current Navigation Tools

Digital moving maps and various electronic assistance tools or an EFB such as ForeFlight [7] and FlightDeck Pro [8] offer improved visualization features, yet their inherent static characteristics frequently overlook the fluid nature of ATCOs' commands. According to Parasuraman and Riley [9], instances of automation being improperly utilized or completely overlooked can arise from discrepancies between the design of the system and the anticipation of its users. This issue is especially significant in taxiway navigation, where pilots face the challenge of time-critical simultaneous navigation (taxiing) and communication tasks. The integration of ATCO clearance into current assistance tools still requires much research and development.

---

[1]The Wizard-of-Oz method is a research technique where participants believe they are interacting with a system, but in reality, the system is operated or partially controlled by a human. This approach allows researchers to simulate automated responses without requiring a fully functional system.

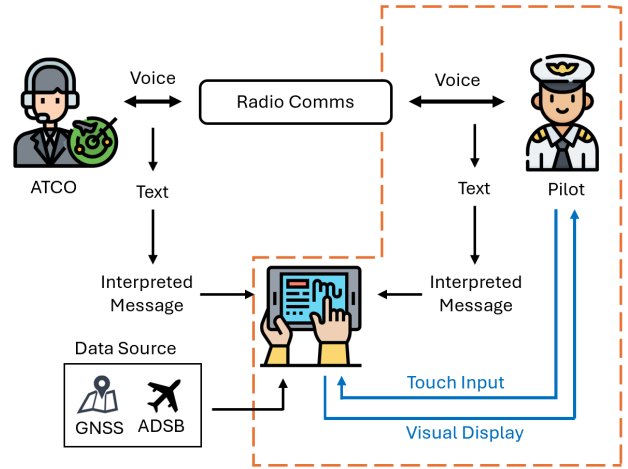

Figure 1. Overview of the proposed taxiway assistance. The system integrates voice inputs from both ATCO and pilot, converting them to text before validating against each other. It also receives positioning and traffic data from Global Navigation Satellite System (GNSS) and Automatic Dependent Surveillance–Broadcast (ADS-B) respectively. The dashed line area represents the scope of this study where different input and output modalities are evaluated with pilot participants.

### B. Input Modalities in Aviation

Recent studies have delved into the effectiveness of different input modalities, such as Paper-Based, Adaptive Keyboard, Map Touch, and Speech-to-Text Recognition. In a surprising twist, Estes et al. [5] revealed that input methods utilizing paper surpassed their digital counterparts in both speed and precision. This discovery calls into question the widely-held belief that technological advancements always improve performance and highlights the necessity of comprehending contextual elements that affect the effectiveness of various modalities. This finding emphasizes the importance of evaluating different input modalities, motivating the analysis of why the paper-based method surpasses its digital counterparts.

### C. Desktop vs Simulation Environment

In aviation research, the choice between desktop-based experiments, high-fidelity simulations, and real-world operations significantly impacts the validity and applicability of findings. Desktop-based experiments offer controlled environments for initial testing but often lack the immersive elements critical for replicating real-world scenarios. High-fidelity simulations, on the other hand, provide a more accurate representation of physical dynamics and complex interactions, closely mimicking real-world conditions. The realism of high-fidelity simulations is particularly valuable in fields like aviation and robotics [10].

Flight simulators vary in fidelity, which refers to how closely they replicate real-world flight conditions. High-fidelity simulators provide full cockpit replication, advanced environmental conditions, and realistic motion feedback, making them suitable for pilot certification and operational training. Low-fidelity simulators include basic desktop-based environments with limited realism, mainly used for procedural training and conceptual learning. In this study, a medium-fidelity flight simulator was selected which aims to improve on Estes et al.'s study [5] done on a desktop, offering an appropriate balance of realism and accessibility, this ensures that it sufficiently replicates real-world taxiing tasks while remaining practical for controlled experimentation.

### D. Research Gaps

Despite significant advancements in aviation technology, gaps remain in our understanding of how automation can be tailored to support pilot-centered operations. Existing studies have primarily focused on static performance metrics, neglecting the dynamic nature of taxiway navigation tasks. Additionally, factors such as clearance complexity and speech recognition accuracy-which directly impact pilot decision-making and system reliability, have received limited attention in the literature.

The main gaps identified by current state-of-the-art research include:

- A lack of qualitative analysis on input modalities done in the existing study of Estes et. al [5] where this study plans to fill the gap.
- In the same study, only one airport was used, which may induce a potential learning effect. Therefore, this study proposes a custom overlay method to reduce the learning effect.
- This study plans to further the work done by Estes et. al [5] by evaluating different input modalities on a medium-fidelity simulator as compared to a desktop.

### III. METHODOLOGY

To address the research gaps identified in this study, we implemented a structured Wizard-of-Oz approach. This approach enabled us to create a realistic and immersive testing environment for users without requiring full automation at this preliminary stage. By simulating automated system responses, researchers could maintain essential oversight and control over critical experimental variables.

### A. Procedures

Once participants arrived at the research facility, they were welcomed with a comprehensive briefing session. This session detailed the aims of the study, explained the experimental arrangements, and reviewed essential safety protocols.

After the briefing, participants made their way to the flight simulator, where they engaged in a 15-minute training session. This introductory phase provided them with an opportunity to familiarize themselves with the simulator's systems and controls. Once the participants felt at ease within the simulator environment, they were equipped with specialized eye-tracking glasses (Pupil Lab Core). A calibration process was performed to ensure precise data collection regarding their visual attention and scanning behavior.

The experiment was designed around four distinct scenarios (see Fig. 2), each using a unique input method: a paper-based, keyboard touch, map touch, and speech-to-text. The scenarios and input methods were systematically counterbalanced among participants to ensure balanced distribution and minimize order effects and prior knowledge influences. Each participant was tasked with executing a taxi clearance activity with the designated input method in every scenario, adhering to the clearance provided by the ATCO. Participants were asked to adopt the "think-aloud" protocol during the experiment, which allows for a more detailed understanding and justifications of their actions and decisions.

Upon completion of each scenario, participants are required to complete a set of subjective assessment questionnaires, and a semi-structured interview. Lastly, after all four scenarios, a catch trial was conducted to further investigate pilots' behavioral attitudes toward the Speech-to-Text method. The catch trials simulate the presence of intentional errors for the speech-to-text input method. It consists of a PowerPoint presentation, simulating the text displayed by the speech-to-text function. While a taxi clearance is verbally given by the ATCO, we manually trigger the word-by-word animation on the screen.

Each participant encountered eight catch trial scenarios varying in clearance length and transcription format. Four scenarios had no errors, while four contained intentional speech-to-text translation errors. Within each condition, two scenarios displayed the full clearance at once, and two transcribed the clearance word by word. This design assessed pilots' responses to errors and different text presentation methods for the speech recognition accuracy.

### B. Software and Hardware for Environment

The simulation software environment was developed using the X-Plane 11 package [11]. Hardware controls consist of a yoke, throttle quadrant, rudder pedals, avionics (consisting of GNS530 Navcom, GTN650 Navcom, GMA350 audio panel, GFC700 Auto-Pilot, two Garmin G5s) and three 55" television monitors for 270-degree creating an immersive environment that replicates real-world flight operations. A depiction of the simulator is presented in Fig. 3.

### C. Participants and Scenarios

The research engaged four licensed pilots (all males) who possessed an average of 1047 total flight hours. The criteria are as follows:

- Holds a current and valid Private Pilot's License (PPL) or above: Ensures that participants possess the necessary foundational knowledge and skills for aviation operations.
- A minimum of 10 flight hours in the preceding year: This criterion verifies that participants are actively flying and current with operational norms and duties.

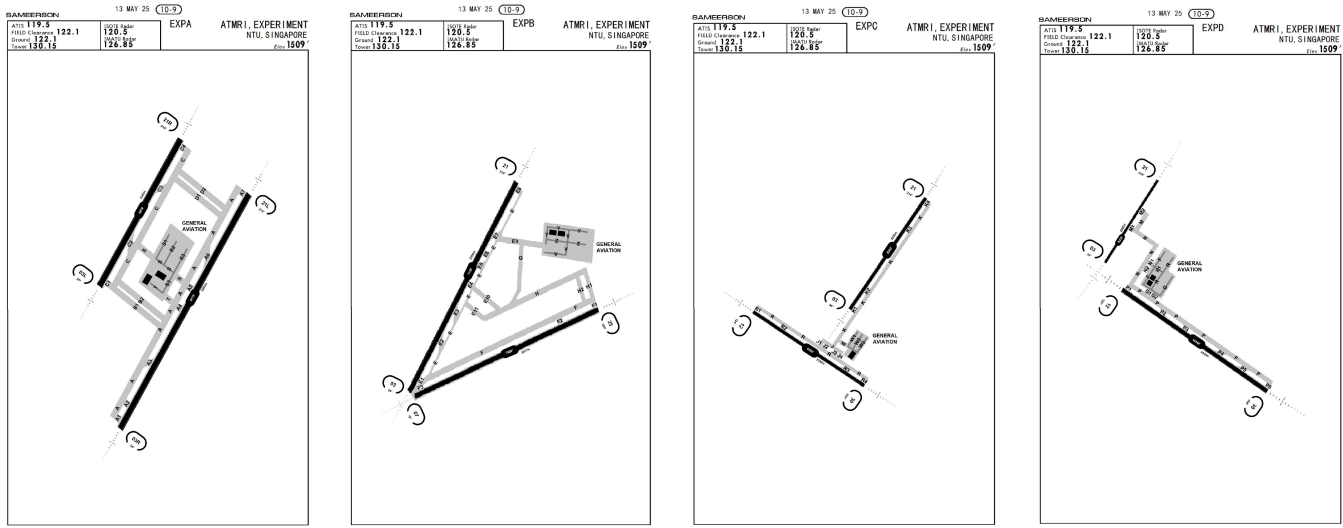

Figure 2. Kiruna Airport (ICAO: ESNQ) which features a custom overlay that balances the complexity of the airport with an additional runway and main taxiway, rendering the airport visually unrecognizable in each scenario.

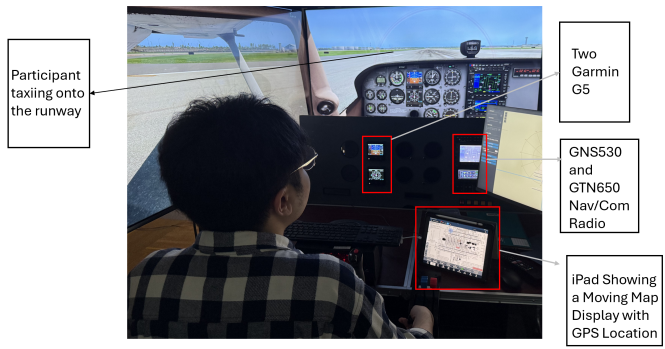

Figure 3. Subject taxiing and aircraft in a medium-fidelity cockpit simulator with 270 degree field of view, two Garmins G5s PFD and HSI, one Garmin GNS530 Nav/com radio, one GTN650 Nav/com radio, with an iPad displaying an airport diagram

The participants were presented with the task of navigating a customized airport map, which was specifically designed to minimize the influence of familiarity biases that might skew the results. This careful approach also took into account potential confounding variables, including each pilot's previous exposure to varied airport configurations. These constructed scenarios were designed to simulate authentic real-world situations that real pilots may encounter during operations. The design aims to ensure that each aspect of the experience was reflective of true-to-life challenges faced in the aviation industry.

The study employed four custom scenarios based on the layout of an existing airport (ICAO code: ESNQ) as shown in Fig. 2. Each scenario incorporated custom overlays that added a fixed number of taxiways and runways to the base layer of ESNQ, creating visually distinct yet comparably complex airport layouts. The complexity of each modified airport was validated pre-experiment by a panel of 10 pilots, who rated the scenarios on a complexity scale of 1–5. All ratings fell between 2 and 3, confirming the consistency of complexity levels across scenarios. Additionally, the custom scenarios were created with reference to FAA AC150-5300 13B [12] and FAA AC150-5340-18H [13] to ensure the logical and feasible design to the modified airport layouts.

*D. Input Methods*

To assess various navigation input modalities, four unique methods were employed: paper-based, keyboard touch, map touch, and speech-to-text. Each approach was crafted to mirror pilot interactions with navigation tools using the Wizard-of-Oz method while ensuring experimental consistency.

The **Paper-Based** method as shown in Fig. 4(a), serves as the baseline for evaluation. This involved participants receiving a conventional airport diagram printed on paper, complemented by a pen to manually copy and trace their intended taxi route. This technique closely mirrored traditional navigation practices that used paper charts, allowing pilots to visualize movements without digital assistance.

The **Keyboard Touch** method, shown in Fig. 4(b) engaged the participants through a user interface presented on a laptop with Touch screen. Here, individuals interacted with the system by selecting taxi routes via keystrokes, emulating the experience of entering taxi clearances using an assistance tool on a tablet or an EFB. The interface provided visual confirmation of the key selected in a textual form and the visuals were shown using the Wizard-of-Oz method by displaying the highlighted route on an airport diagram to the participant. This is to simulate the illumination of the chosen taxiway path which offers a structured yet digitally enhanced substitute for conventional paper navigational methods.

In the **Map Touch** method, shown in Fig. 4(c) participants were provided with an airport diagram on a piece of paper,

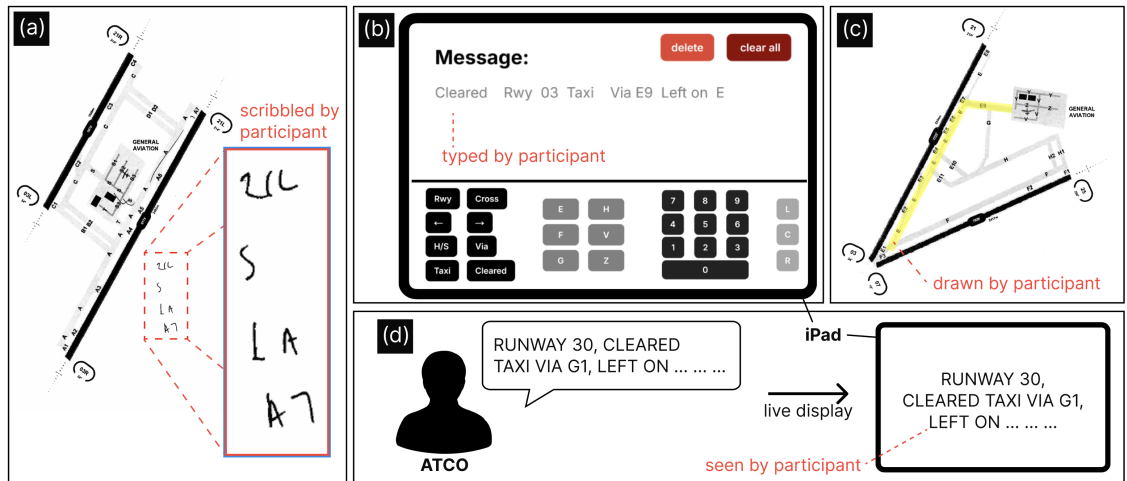

Figure 4. Overview of input methods tested: (a) **Paper-Based**: Participants scribbled clearances using a pen; (b) **Keyboard Touch**: Participants tapped buttons on the lower half of the iPad, with text appearing at the top; (c) **Map Touch**: Participants highlighted the cleared taxi route; (d) **Speech-to-Text**: Simulated live transcription as the ATCO verbalizes the clearance.

where they utilized a highlighter tool to delineate their taxi route. This method replicated the use of touchscreen-based navigation aids where users can touch to highlight a digital representation of the airport taxi diagram.

The **Speech-to-Text** method as shown in Fig. 4(d) was executed using a Wizard-of-Oz framework, received a verbal taxi clearance from the ATCO and the clearance is transcribed into text displayed on a laptop screen, simulating an interaction with an automated speech recognition system rather than employing an authentic speech-to-text mechanism. A pre-prepared taxi clearance text was displayed manually on a screen via Microsoft PowerPoint to emulate the system. This strategy facilitated controlled assessments of the speech modality while minimizing the unpredictability introduced by real-time speech recognition challenges.

By comparing these four distinct input methods, the study aimed to investigate the effectiveness in terms of task completion time, usability, and workload linked to each modality, thereby yielding valuable insights into their potential applicability in future aviation contexts.

*E. Study Design*

This section outlines the key variables and post experiment questions examined in the study. The following sections detail the independent, dependent variables and the questions asked for the evaluation after each run.

Independent variables:
1. Input Modality: The four input modalities (paper-based, keyboard touch, map touch, and speech-to-text) were tested, with the paper-based method serving as the baseline
2. Speech Recognition Accuracy: Simulates either perfect or error-prone speech recognition systems

Dependent variables:
1. Task Completion Time: defined as the interval from ATCO clearance completion, to the start of taxi. This measure is comparable to that of Estes et al. [5] which used the time from when clearance was issued to when the pilot indicated 'ready to taxi'.
2. Subjective Workload: was measured using NASA-TLX (Task Load Index) [14], which evaluates workload across six dimensions: mental demand, physical demand, temporal demand, performance, effort, and frustration. Each dimension was rated on a 7-point scale,
3. Usability: was measured using System Usability Scale (SUS) [15], a widely used metric for evaluating user experience. It consists of 10 standardized questions, each rated on a 5-point Likert scale (from "Strongly Disagree" to "Strongly Agree"), yielding a composite score that reflects the ease of use and intuitiveness of each method.

Beyond these structured assessments, participants were also asked a series of open-ended questions at the end of the whole session to gain deeper insights into their experiences. Participants were asked open-ended questions about their experiences and challenges in ATC communication, suggestions for improvements to a taxiway assistance, and rankings of the input methods based on personal preferences. Participants were also encouraged to reflect on potential problems associated with each method in operational environments.

The collected data was analyzed using both quantitative and qualitative approaches. Repeated measures ANOVA was conducted to analyze the system usability on the SUS score and as well as the on the Task Completion Time, using each of the different input modality. Friedman's test (a non-parametric alternative to the one-way ANOVA) was used to analyze the NASA-TLX score. Additionally, qualitative feedback from open-ended responses was examined to uncover recurring themes and user preferences. This comprehensive analysis provided critical insights into the effectiveness and feasibility of each input method, informing future improvements in taxiway assistance and ATCO communication systems.

## IV. RESULTS

### A. Quantitative Results

Given the small sample size (n=4), inferential statistics were not expected to demonstrate statistical significance and therefore, the descriptive overview of data is presented.

First, the task completion time of each participant is ranked from the fastest to the slowest, which is represented '1' and '4' respectively. The overall ranking is shown as Table I. Three out of four participants took the longest time to complete the task using the baseline paper-based method whereas the speech-to-text method was the fastest for half of the participants.

TABLE I. RANKED TASK COMPLETION TIME

| Participant | Paper-Based | Keyboard Touch | Map Touch | Speech-to-Text |
|---|---|---|---|---|
| P1 | 4 | 3 | 1 | 2 |
| P2 | 4 | 1 | 2 | 4 |
| P3 | 3 | 4 | 2 | 1 |
| P4 | 4 | 2 | 3 | 1 |

The NASA-TLX scores showed no significant difference amongst the four inputs. The paper-based method was used as the baseline to compare the other three modalities. The workload for the map touch method was 2.80±0.48, which is lower than that of the speech-to-text method (2.96±1.22). The SUS scores were analyzed, and no significant differences was found among the four input methods, with the highest being 91.25±10.31 (map touch) and the lowest being speech-to-text (72.50±31.56). The SUS scores of all four proposed formats are "acceptable" based on the acceptability range as shown in Fig.5 [15]. The map touch is the only method with a higher SUS score than the paper-based (baseline). Both the keyboard touch and speech-to-text methods had a usability score lower than that of the paper-based method.

TABLE II. SUMMARY OF MEAN±STANDARD DEVIATION ACROSS THE FOUR INPUT METHODS

| Measure | Paper-Based | Keyboard Touch | Map Touch | Speech-to-Text |
|---|---|---|---|---|
| NASA-TLX | 3.67 ± 0.99 | 3.00 ± 0.76 | 2.80 ± 0.48 | 2.96 ± 1.22 |
| SUS | 88.30 ± 3.75 | 74.38 ± 27.94 | 91.25 ± 10.31 | 72.50 ± 31.56 |
| Response Time | 7.24 ± 7.48 | 2.72 ± 1.45 | 8.76 ± 8.37 | 1.60 ± 0.64 |

### B. Qualitative Results

From the open-ended questions, the user preference is inconclusive as presented in Table III. However, in the open-ended questions, Three out of four participants answered having the "follow-the-green" or a system that allows for progressive taxi as a preferred feature they would like to have in any taxiway assistance application. Three out of four participants answered the automated highlight function as the most critical feature.

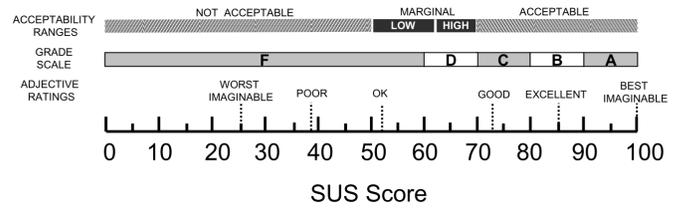

Figure 5. The means of SUS scores of the four inputs range from 72.50 to 91.25, all of which are "acceptable" based on the acceptability range [15].

The open-ended questions also revealed multiple challenges currently faced by the participants as follows:
- Difficulty understanding controllers with strong accents
- Overwhelming information given in a single clearance
- Muffled radio calls or overlapping transmissions from other pilots

From the qualitative results gathered, the results are further elaborated and discussed in the Discussion section.

TABLE III. USER PREFERRED METHOD OF INPUT MODALITY

| Measure | Paper-Based | Keyboard Touch | Map Touch | Speech-to-Text |
|---|---|---|---|---|
| No. of preferred | 1 | 1 | 1 | 1 |

### C. Eye Tracking Results

The heatmap aggregates fixation frequency and duration to illustrate the spatial distribution of gaze points over time, reflecting how long pilots spend processing information in different areas of interest (AOIs) [16]. Two AOIs were defined to analyze gaze distribution and attention allocation: the cockpit and the view outside the window, and below the cockpit, representing external navigation and situational awareness, and below the cockpit, encompassing instruments, controls, and reference materials used during operations.

As shown in Fig.6, for methods involving paper charts and map touch navigation—which are closer to traditional pilot training—fixations are concentrated on the cockpit, the forward view of the taxiway, and the paper/ map. This indicates that during taxiing, pilots frequently shift their gaze between these three areas. In contrast, when using the speech and keyboard methods, pilots must cross-reference the map with on-screen messages to confirm the taxi route. This process, as indicated by the gaze pattern forming a saccade within 3 seconds (see Fig.7, results in more frequent downward scanning and head movements.

## V. DISCUSSION

The absence of statistically significant differences in our quantitative data is likely attributed to the limited sample size. Nonetheless, the qualitative feedback offers valuable insights and suggests several avenues for future research. In particular, participants expressed diverse preferences regarding input methods. A summary of the Pros and Cons of each input modality can be found in Table V. Next, each input method was examined individually, outlining the rationale behind participants' positive

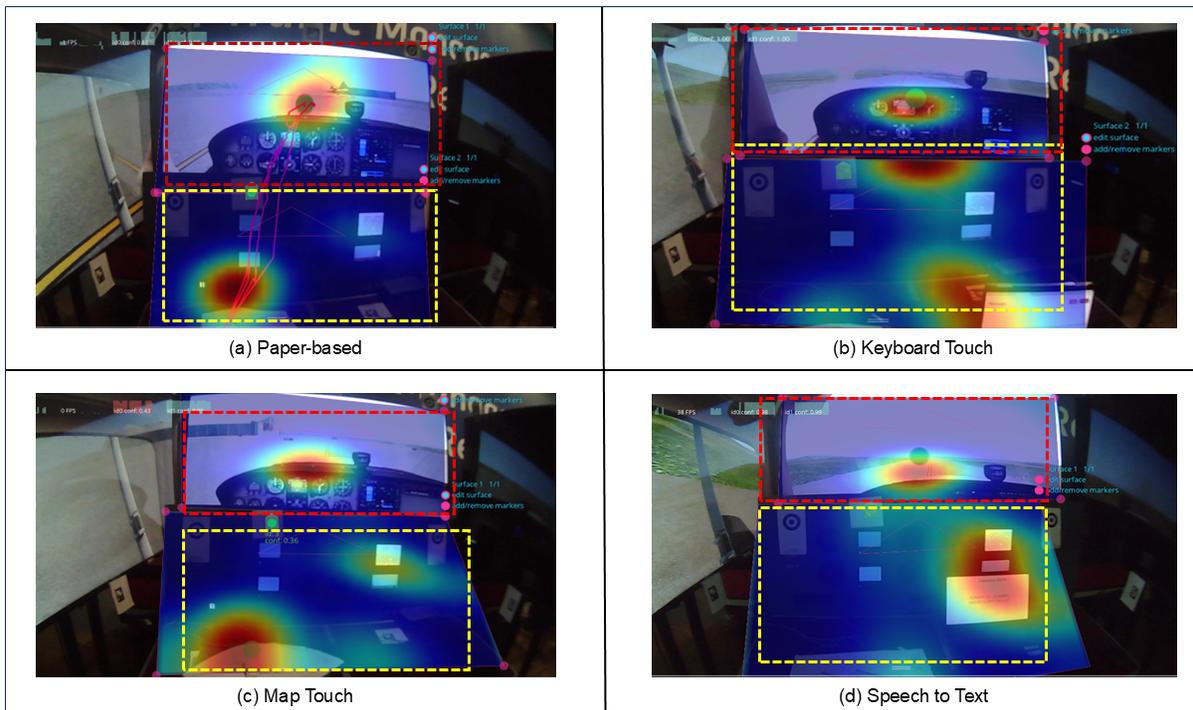

Figure 6. Heatmaps of gaze patterns across four input methods: The red and yellow dashed boxes represent AOI1 and AOI2, respectively. These visualizations show that pilots frequently shift their gaze between the cockpit, the forward view of the taxiway, and the paper or map when using paper and map inputs.

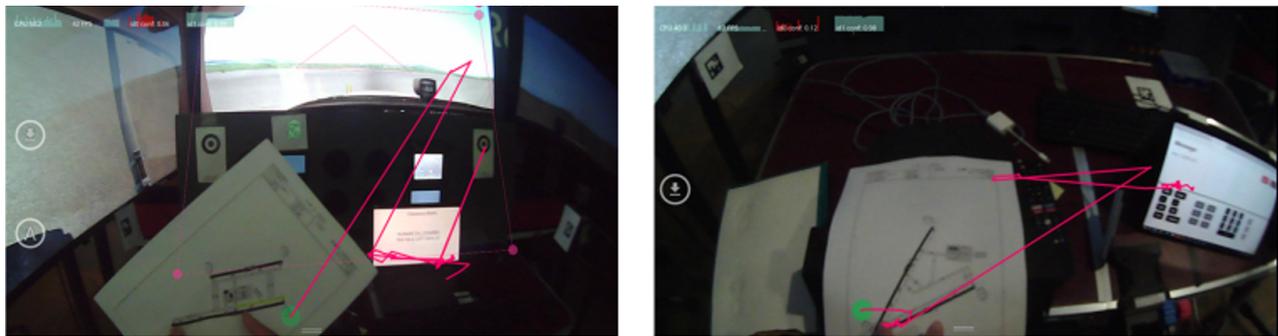

Figure 7. Saccade patterns within 3 seconds: Gaze trajectories of using Speech-to-Text (left) and Keyboard Touch (right) input methods as taxiway navigation references, observed during pilots looking downward.

and negative evaluations followed the the discussion on the task completion time.

The **Paper-Based** method elicited polarized responses. P2 ranked it the most preferred method, citing its alignment with long-term habits developed during training and operational practice. In contrast, the remaining three participants ranked the paper-based method as second-to-last or last. They expressed concerns about the risk of manual transcription errors, such as mishearing or miscopying clearances—a critical risk in time-sensitive scenarios. These observations are consistent with the findings of Harter et al. [17] who reported high incidence of documentation errors in handwritten processes. P2's preference likely stems from the mere-exposure effect, where repeated use of paper-based systems fosters psychological comfort, even if sub-optimal. Although the familiarity of the paper-based method is advantageous, its reliance on human precision introduces potential reliability trade-offs. Future studies could consider increasing the number of trials and incorporating more complex clearance scenarios to determine whether the paper-based method remains preferable under more realistic conditions.

The **Keyboard Touch** method was praised for its perceived efficiency and adaptability. P1 ranked it first, while both P2 and P3 ranked it second. Participants appreciated its tactile feedback and straightforward design, likening it to a "digital" version of the paper-based method that requires minimal re-training. However, participants raised concerns about error recovery and scalability. For error recovery, P3 noted that when addressing a mis-click, it was challenging to follow with newer instructions from the ATCO. For scalability, P2 added that complex airports with numerous taxiways (labelled A–Z) might overwhelm users during visual searches, potentially increasing the risk of mis-clicking, and thereby delaying the overall clearance copy process. This observation aligns with Hick's Law [18], which posits that reaction time increases with the number of choices. Such limitations suggest that while our keyboard touch method may excel in smaller airport settings, it could be less effective in larger, more complex settings. Future iterations may benefit from leveraging the standard QWERTY keyboard layout to spatially arrange the letters A to Z, and to adopt context-aware keys [19] (e.g., activating only relevant taxiway keys based on airport maps. These adaptations could help focus the user's attention on the most relevant keys while capitalizing on their prior familiarity with the standard layout, thereby enhancing kinesthetic ease of use.

The **Map Touch** method was lauded for enhancing spatial cognition. P3 ranked it as first, emphasizing that visualizing taxi paths reduced cognitive workload, and we reasoned because of leveraging visual working memory [20]. P3 noted that the straightforward design—where users simply identify the taxiway connecting to the initially highlighted path—provided clear visual affirmation for subsequent readback and taxi operations. In particular, P3 remarked that "a picture is worth a thousand words", highlighting how the Map method saved time and effort compared to manually writing out clearances (as in the paper-based method), which still conveys the intended message. However, its effectiveness may be compromised if errors occur, and appear to depend on user familiarity. Similar to the keyboard method, P3 and P4 indicated that in scenarios involving very long clearances, a mistake could lead to challenges in refocusing on later instructions issued by the ATCO. P4 struggled while navigating unfamiliar airports and this revealed a critical dependency on prior knowledge, suggesting that future should consider customizable zoom and rotation features to better accommodate varying levels of user familiarity in the environment. Additionally, P1's critique of ambiguous directional cues (e.g., "hold short A1" vs. "hold short runway 03R on A1") points to a need for future research to integrate standardized symbols (e.g., arrow) to ensure unambiguous communication.

The **Speech-to-Text** method received mixed feedback. P4 praised its dual text-and-visual output for enhancing situational awareness, noting it allowed them to "monitor for errors rather than transcribe". However, distrust in automation was pervasive: P1, P2, and P3 expressed skepticism about transcription accuracy, echoing Parasuraman & Riley's [9] analysis that suggests if users perceive automation as imperfect or unreliable, they may be inclined to disuse it, even in cases where the automation performs better than manual control. P1 cited the cognitive workload of validating outputs, stating, "if the system is wrong, there's no quick way to correct it." P2 also added that transcription errors could cascade if users "forget where the mistake occurred" during faster clearances. Furthermore based on the catch-trial study, display rate may influence pilot's error detection. For instance, when the clearance is presented one word at a time, participants could spot 75% of the errors, as compared to presenting one clearance block (all the words) at a time, resulting in a reduced 50% of error detection. This reduction could be explained by theories of incremental language processing combined with the "good-enough" processing perspective. In incremental processing, readers build up a sentence's meaning one word at a time, actively integrating each new word with what came before. When the clearance is revealed word-by-word, this forces a serial, detailed analysis [21], where any anomaly or error is immediately noticeable because it is not aligned with what they are hearing live. By contrast, when the entire sentence is visible at once, readers are more likely to rely on a "good-enough" or heuristic approach—grasping the overall gist rather than scrutinizing each word [22]. This global processing can sometimes allow local errors to go undetected. To mitigate distrust, future systems could integrate (1) real-time correction features (e.g., touchscreen edits during pauses) and (2) confidence scores for transcribed segments, fostering transparency. Hybrid interfaces pairing speech input with tactile validation (e.g., tapping to confirm taxiways) might balance automation benefits with user control.

Participants' preferences reflect trade-offs between familiarity, precision, and workload in taxiway navigation. While paper-based and keyboard touch methods align with pilots' ingrained habits, map touch and speech-to-text interfaces offer potential efficiency gains—but at the cost of trust and adaptability. These findings challenge assumptions about digital superiority, emphasizing that automation must be designed with a pilot-centered approach to enhance usability. The distrust in speech-to-text systems highlights the need for real-time error correction, confidence indicators, and redundancy measures to improve adoption. Meanwhile, the strong usability of map-based interactions suggests that digital solutions should leverage spatial cognition principles to improve clearance comprehension and execution.

Future research should focus on hybrid systems that merge the benefits of traditional and digital methods. A keyboard interface integrated with a map-based visual validation system could provide both input flexibility and situational awareness. Additionally, longitudinal studies are needed to assess whether increased exposure to automation reduces distrust over time, addressing a critical barrier to adoption in aviation operations.

In terms of the task completion time, the Baseline paper-based method was found to be the slowest in three out of four participants, which does not align with Estes et. al.'s study. In the study, the authors found that the paper-based method was the fastest, followed by the Keyboard method. They also noted that if the Speech recognition is error-free, the Speech method would be the fastest. Our results shows otherwise, where only half of the participant showed the fastest task completion time using the speech-to-text method followed by the map touch method. In the paper-based method, P3's Keyboard result is the slowest. A probable cause could be due to his age (10 years above the mean age of participants) making him less tech savvy which may affect his speed in adapting to new technology.

TABLE IV. COMPARISON OF CURRENT SYSTEMS VS. FINDINGS FROM THIS STUDY

| Feature | Current Systems | Findings from This Study |
|---|---|---|
| **Pilot Efficiency** | Assumed to improve response times | Paper-based methods still competitive in accuracy and speed |
| **Speech-to-Text** | Under development, but expected to enhance automation | Low pilot trust due to error concerns; requires real-time correction |
| **Cognitive Workload** | Expected to reduce workload | Varies by input method; map touch reduces load, speech increases it |
| **Error Mitigation** | Some automation, but limited real-time correction | Manual input methods allow direct corrections; automation needs trust-building features |
| **System Adaptability** | Limited adaptation to real-time ATC changes | Hybrid approaches (map + keyboard) may provide better flexibility |

TABLE V. SUMMARY OF PROS AND CONS OF EACH INPUT MODALITY

| | **Paper-Based** | **Keyboard** | **Map Touch** | **Speech-to-Text** |
|---|---|---|---|---|
| Pros | • Trained conventionally to use paper and pen | • Enhanced version of paper method, does not require much training<br>• Facilitated user control | • Enhanced clarity of cleared path with visualization<br>• Added level of affirmation when selecting the taxiway | • Do not need to do anything physically, only monitor if the clearance is correct |
| Cons | • Possibility of making a human error | • Difficulty in locating specific key in a complex airport<br>• Could potentially press wrong<br>• Difficulty with following the ATC's clearance when correction is required. | • Hard to locate specific taxiway if unfamiliar<br>• Does not show direction<br>• Difficulty with following the ATC's clearance when correction is required. | • Multiple participants do not trust the automation |

## VI. LIMITATION AND FUTURE WORK

While this study provides insights into the impact of different input modalities on pilot taxiway navigation, certain limitations must be acknowledged:

1. Wizard-of-Oz – This study used a Wizard-of-Oz method to simulate different input modality instead of a functioning system. While this ensured consistent system responses and eliminated variations due to recognition errors, it does not fully account for real-world challenges such as system error/failure (e.g error in speech recognition). Future research should incorporate a working prototype of the system to better evaluate their practical reliability and usability.
2. Limited Sample Size – The study was conducted with a specific subset of licensed pilots who met a pre-determined criteria. While this ensures that all participants were active and experienced in flight operations, the sample size may limit generality of the findings to broader pilot populations, including those with varying levels of experience, different operational backgrounds, or exposure to different taxiway navigation systems.
3. Static Weather and Traffic Conditions – The experimental setup did not account for dynamic real-world variables such as adverse weather conditions (e.g., fog, rain, or snow) or high-density airport traffic scenarios. These factors can significantly influence pilot decision-making, situational awareness, and cognitive workload. Future studies should incorporate simulated environmental variations to assess how different input modalities perform under more challenging operational conditions.

Building upon the findings of this study, future work will focus on evaluating pilot performance under low visibility conditions, where reliance on an advanced taxiway assistance becomes essential for safe and efficient taxi operations. Unlike the current study, which explored different input modalities, the next phase of research will investigate how pilots interact with a fully digital taxiway assistance in environments where out-of-window visibility is significantly restricted. A key area of exploration will be eye-tracking analysis, including heatmap visualization and saccadic patterns, to understand how pilots allocate visual attention when navigating solely through an on-screen taxiway assistance. By examining these gaze behaviors, we aim to determine whether pilots can effectively interpret and act on digital information to taxi the aircraft to the designated clearance position provided by ATCO.

The enhanced taxiway assistance will integrate several key features to facilitate low-visibility taxiing, including:
- Live aircraft positioning via GNSS.
- Real-time traffic data displaying nearby aircraft and ground vehicles.
- Visualized airport map with highlighted cleared taxi routes and dynamic markers for hold-short positions (e.g., red lines to indicate stop points).
- Warning and advisory messages for situational awareness.
- Automated speech-to-text transcription of taxi clearances, with an editable function to correct potential recognition errors.

This future study will be conducted in a simulated environment to validate the feasibility of solely relying on a digital taxiway assistance system for ground movement in low-visibility conditions. Additionally, the system's effectiveness will be evaluated in a highly complex airport environment, where taxiing presents greater challenges due to intricate layouts and increased traffic density.

## VII. CONCLUSION

This study provides a comprehensive evaluation of different input modalities on pilot performance in taxiway navigation, addressing key limitations in prior research and challenging the assumption that digital solutions always outperform manual methods. By conducting experiments in a medium-fidelity simulator with four customized airport layouts, we mitigated learning effects and ensured a more ecologically valid assessment.

They key findings are summarized as follows:

1. Traditional paper-based methods remain competitive, despite the rise of automation
2. Map touch interfaces improve situational awareness but require intuitive error handling
3. Speech-to-text systems suffer from low pilot trust, highlighting the need for hybrid solutions
4. Future taxiway assistance systems must prioritize real-time adaptability and cognitive load reduction

The results indicate that traditional paper-based methods, despite being low-tech, remain highly effective in specific contexts. However, digital alternatives, such as speech-to-text input, hold significant promise if integrated with improved recognition accuracy and adaptive automation. Additionally, the use of counterbalancing techniques ensured that order effects were minimized, strengthening the validity of our results. Future research should focus on refining automation strategies, particularly in enhancing speech recognition reliability and dynamic system adaptability. Further evaluations in operational environments could provide deeper insights into the practical implementation of these findings. Ultimately, this study contributes to the ongoing efforts to enhance airport ground operations by improving pilot navigation tools and fostering trust in human-machine collaboration.

These findings highlight that while digital taxiway navigation tools offer efficiency gains, their effectiveness depends on pilot trust, interface adaptability, and cognitive workload considerations. The study reinforces the need for hybrid systems that integrate the familiarity of traditional methods with the flexibility of digital interfaces to optimize both accuracy and usability. Future taxiway assistance should prioritize pilot-centered automation, incorporating real-time feedback, error correction mechanisms, and intuitive visual aids to enhance situational awareness. As aviation technology advances, striking the right balance between automation and human adaptability will be crucial in reducing pilot deviations and improving ground navigation safety.

## ACKNOWLEDGMENT

This research is supported by the National Research Foundation, Singapore, and the Civil Aviation Authority of Singapore, under the Aviation Transformation Programme. Any opinions, findings and conclusions or recommendations expressed in this material are those of the author(s) and do not reflect the views of National Research Foundation, Singapore and the Civil Aviation Authority of Singapore. The authors thank Louis Hoo Xuan Tao, Thanh Danh Le, Hasnain Ali and Dong Liang for their support. This research has also been reviewed and approved by Nanyang Technological University's (NTU) Institutional Review Board (IRB-2024-1105). Some portions of this paper were refined with the assistance of ChatGPT, an AI language model developed by OpenAI. The final content was reviewed and edited by the authors.